\documentstyle[preprint,aps]{revtex}
\draft
\begin{document}

\title{Diffusion of electrons in two-dimensional disordered
symplectic systems}
\author{Tohru Kawarabayashi}
\address{Institute for Solid State Physics, University of Tokyo,
Roppongi, Minato-ku, Tokyo 106, Japan}
\author{Tomi Ohtsuki}
\address{Department of Physics, Sophia University,
Kioi-cho 7-1, Chiyoda-ku, Tokyo 102, Japan}

\date{\today}
\maketitle

\begin{abstract}
Diffusion of electrons in two-dimensional disordered systems with
spin-orbit interactions is investigated
numerically. Asymptotic behaviors of the
second moment of the wave packet and of the temporal auto-correlation
function are examined.
At the critical point, the auto-correlation function exhibits
the power-law decay with a non-conventional exponent $\alpha$
which is related to the fractal structure
in the energy spectrum and in the wave functions.
In the metallic regime,
the present results imply that transport properties
can be described by the diffusion equation for normal metals.

\end{abstract}


\narrowtext

The Anderson transition in disordered systems is usually
classified into three universality classes,
the orthogonal,
the unitary and the symplectic\cite{Dyson}. The classification
depends only on the symmetry of the system.
In the case where the Hamiltonian is represented by a real and
symmetric matrix, the system belongs to the orthogonal class.
Systems having no time reversal symmetry belong to the unitary
class.  The system with a spin-dependent interaction having the
time reversal invariance  belongs to the symplectic class.

In two dimensions, these three universality classes show quite
different critical behaviors. In the case of the orthogonal class,
all the states are considered to be localized.
In the case of the system
in a magnetic field, which is a typical example of
the unitary class,
all the states are again localized except at the centers of the
Landau bands. Anyway, in two dimensions,
these two classes do not show the Anderson transition. In contrast,
the symplectic class is
known to exhibit the Anderson transition in two-dimensional
systems\cite{EZ,Ando,E,FABHRSWW,Fa}.

A typical system which belongs to
the symplectic universality class is a disordered
system with spin-orbit interactions.
It has been proved that
in two dimensions,
such a system has localized states if the disorder is strong enough
compared to the spin-orbit coupling\cite{BGMS}.
Numerical and analytical studies including the
finite-size scaling method \cite{EZ,Ando,E,FABHRSWW,Fa}
and the renormalization group theory
\cite{HLN,Hikami,Grempel,Kawabata} have been made on
the critical behavior of the Anderson transition
in two-dimensions.
However, several important issues still remain to be clarified.
In particular, the transport properties in the metallic regime
appear to be peculiar. First,
a direct application of
the scaling theory leads to a result that
the conductivity diverges logarithmically
\cite{HLN,Kawabata}.
The infinite conductivity
in the presence of
the scattering by the impurities does not seem to be physically
acceptable.
Second, the scaling function
in the metallic regime, obtained by the
finite-size scaling method\cite{Ando,E}, is different from
what we expect from the theory developed by MacKinnon and Kramer
\cite{MK} for
the orthogonal case.
These facts imply that the metal-insulator transition
in two dimensions
may have some complex structures which have to be explored further.
The validity of the one-parameter
scaling hypothesis in such a case, for instance,
has already been discussed \cite{E,Markos}.
It is thus an interesting problem to clarify the transport
properties in the metallic regime and at the critical
point of two-dimensional systems.

In this letter, we examine the asymptotic behaviors of the
temporal auto-correlation function and of the second moment of
the wave packets in the two-dimensional disordered system with
spin-orbit interactions.
Since
the Anderson transition in two dimensions
occurs only  in the case of symplectic ensemble
the emphasis is placed
on the behaviors of these quantities
in the metallic regime and at the critical point.
The second moment of the wave packets and the temporal
auto-correlation function are defined by
\begin{equation}
 \langle r^2(t) \rangle = \int {\rm d}{\mbox{\boldmath $r$}}
 r^2 |\psi({\mbox{\boldmath $r$}},t)|^2
\end{equation}
and by
\begin{equation}
 C(t) = \frac{1}{t}\int_0^t P(t'){\rm d}t' , \quad
 P(t) \equiv |\langle \psi(t) | \psi(0) \rangle|^2 ,
\end{equation}
respectively.
Here $\psi(t)$ and $P(t)$ denote the wave function at time $t$ and
the overlap function between the initial state $|\psi(0) \rangle$ and
the state $|\psi(t)\rangle$ at time $t$, respectively.

In the metallic regime,
the second moment in $d$ dimensions is usually expected to
grow in proportion to time $t$ as
\begin{equation}
 \langle r^2 \rangle \sim 2dDt ,
 \label{3}
\end{equation}
where the diffusion coefficient is denoted by $D$.
The diffusion coefficient $D$ is related to the conductivity $\sigma$
by the Einstein relation
 $\sigma = e^2 D \rho$\cite{KTH},
where $\rho$ is the density of states at the Fermi energy.

The auto-correlation function $C(t)$, on the other hand,
is expected to
decay in
proportion to $t^{-\alpha}$
in the metallic regime and at the critical point.
It has been shown that the exponent
$\alpha$ is related to the correlation dimension $\tilde{D_2}$
of the energy
spectrum as $\alpha = \tilde{D_2}$ \cite{KPG,HS}.
It has been further demonstrated that the correlation
dimension $\tilde{D}_2$
is also related to the generalized dimensionality $D_2$ of the
eigenfunctions as $\tilde{D}_2 = D_2 /2$\cite{HS,Schweitzer}.
We can thus obtain information about the fractal structure in energy
spectrum and in the eigenfunctions by evaluating the exponent
$\alpha$.

If the time-evolution of the wave packet
can be described by the diffusion equation,
the auto-correlation function would decay
in proportion to $t^{-d/2}$,
where $d$ denotes the dimensionality of the system.
Thus the conventional value of $\alpha$ is 1
in two dimensions, and this
value of $\alpha$ implies that $\tilde{D_2} = 1$ and $D_2 = d =2$,
indicating the absence of any fractal structure in this system.
The deviation of $\alpha$ from 1 can therefore
be regarded as evidence for a
fractal structure in energy spectrum and in the eigenfunctions.

In the localized regime, the wave packet does not diffuse.
Hence, both the second moment and
the auto-correlation function  have a finite
and nonzero value
in the limit $t \rightarrow \infty$.

The model we adopt in the present letter
is the Ando model \cite{Ando} described by the
Hamiltonian
\begin{equation}
 H = \sum_{i,\sigma} \varepsilon_i C_{i\sigma}^{\dagger} C_{i\sigma}
 - \sum_{<i,j>,\sigma,\sigma'}
 V_{i\sigma;j\sigma'} C_{i\sigma}^{\dagger}
 C_{j\sigma'}
\end{equation}
with
\begin{equation}
 V_{i;i+\hat{x}} = \left( \begin{array}{cc}
              V_1 & V_2 \\
              -V_2 & V_1
              \end{array} \right)
 , \quad
 V_{i;i+\hat{y}} = \left( \begin{array}{cc}
              V_1 & -{\rm i}V_2 \\
              -{\rm i}V_2 & V_1
              \end{array} \right)   ,
\end{equation}
where $C_{i\sigma}^{\dagger}(C_{i\sigma})$ denotes a creation
(annihilation) operator of an electron on the site $i$ with
spin $\sigma$ and $\varepsilon_i$ denotes the site-diagonal
potential distributed independently. The probability
distribution of the site-potential is uniform
in the range $[-W/2,W/2]$. Here, the unit vector of the $x(y)-$%
direction is denoted by $\hat{x}(\hat{y})$. All length-scales
are measured in units of the lattice spacing $a$.
The effect of the spin-orbit interaction is taken into account
by the off-diagonal matrix element $V_2$ which is
chosen such that the relation $V_2/V = 0.5$ holds,
where $V\equiv \sqrt{V_1^2 + V_2^2}$.
For the states at the center of the band $(E=0)$, the critical
point of this model
has been estimated as $W_c = 5.74 V$ by the finite-size
scaling method \cite{Ando,FABHRSWW}.
We confine ourselves in this letter to the case of the
band center $(E=0)$ for simplicity.

To solve the time-dependent Schr\"{o}dinger equation
numerically,
we have used the method based on the
exponential product formulas\cite{Suzuki,DeRaedt,KO}.
One of the advantages of our method is that we can treat
much larger systems than those treated
in the exact diagonalization method.
It should be emphasized that in the present approach,
the time-reversal symmetry, which characterizes the
symplectic ensemble, is exactly preserved if the
symmetric decomposition formula\cite{Suzuki} is adopted.

The essence of the present method is to evaluate the time-evolution
operator $U(t) = \exp(-{\rm i}Ht/\hbar)$
by using the decomposition formula
for exponential operators \cite{Suzuki}.
The $n$-th order decomposition $U_n$
satisfies the condition
\begin{equation}
 U(\delta t) = U_n(\delta t) + O(\delta t^{n+1}) ,
 \label{a1}
\end{equation}
where $\delta t$ denotes the time step of the simulation.
We have adopted the forth-order decomposition, as in the
previous paper \cite{KO}, given by
\begin{equation}
 U_4 = U_2(-{\rm i}pt/\hbar)U_2(-{\rm i}(1-2p)t/\hbar)
 U_2(-{\rm i}pt/\hbar)
 \label{ourdec}
\end{equation}
with
$$
 U_2(x) \equiv {\rm e}^{xH_1/2}\cdots{\rm e}^{xH_{q-1}/2}
 {\rm e}^{xH_q}
 {\rm e}^{xH_{q-1}/2}\cdots{\rm e}^{xH_1/2},
$$
where $H=H_1 + \cdots + H_q$ and $p=(2-\sqrt[3]{2})^{-1}$.
The decomposition $U_n$ is called symmetric
if it satisfies the condition
that $U_n(t)U_n(-t) = 1$ \cite{Suzuki}.
It is easy to verify that the decomposition $U_4$
indeed satisfies  this condition.
It can be shown that the symmetric decomposition $U_4$
has the property
\begin{equation}
 \Theta U_4(t)^{\dagger} \Theta^{-1} = U_4(t),
\end{equation}
provided that the Hamiltonians $H_1,\ldots ,H_q$ are
time reversal invariant,
where $\Theta$ denotes the time-reversal operator\cite{Sakurai}.
Consequently, we have relationships for the
time-evolution of the states as
 $\langle i; + | U_4(t)| j; - \rangle =
- \langle j; + | U_4(t) | i; - \rangle $
and
 $ \langle i; + | U_4(t) | j; + \rangle   =
 \langle j; - | U_4(t) | i; - \rangle $.
Here $|i;\pm \rangle$ represent the states
of the spin up$(+)$ and down$(-)$ electrons at the site $i$.
It should be emphasized that preserving the
symmetry of the system in the numerical approach is quite
important because the properties of two-dimensional disordered
systems strongly depend
on their symmetries.

Numerical calculations are performed in systems
whose sizes are $399 \times 399 \sim 999 \times 999$
and for $5\sim 10$ realizations of potentials.
The
initial state is built by diagonalizing
a system of radius $L=21$ located
at the center of the whole system\cite{KO,DR}.
The time step is chosen as $\delta t = 0.2 \hbar / V_1$.

Let us first discuss the temporal auto-correlation function.
The logarithm of the auto-correlation functions $\log_{10} C(t)$
for the localized regime $(W=9V)$, the extended regime $(W=1V)$ and
the critical point $(W=5.74V)$ are shown in Fig. 1.
For $W=1V$, we find that
the auto-correlation function exhibits the power-law decay and
its exponent is evaluated to be 1. This implies that the
system does not have any fractal structure here and that the
diffusion of electrons in this region can be
described by the diffusion equation.
At the critical point $W=W_c$, the auto-correlation function
again
shows the power-law decay.
However, in contrast to the case of $W=1V$, the exponent
takes a value different from
1, reflecting the fractal structure at the critical point
\cite{Schweitzer,CDEN}
in energy spectrum and in eigenfunctions.
In the localized regime $(W=9V)$, the auto-correlation function
is likely to have a nonzero value
in the limit $t\rightarrow \infty$,
as expected.

At the critical point,
the value of
the exponent $\alpha$ is estimated to be
$\alpha = 0.84 \pm 0.03$.
By assuming the relation $D_2 = 2\alpha$\cite{HS,Schweitzer},
the generalized dimension
$D_2 $ of the wave function
is then obtained as  $D_2 = 1.68 \pm 0.06$.
Our estimate of the exponent
$\alpha$ agrees with the value of the correlation dimension
$\tilde{D}_2 (=0.83 \pm 0.03)$
obtained by the exact diagonalization method \cite{Schweitzer}
and is consistent with the
relationship $\alpha = \tilde{D}_2$ \cite{KPG}.

Furthermore, we have made simulations for other types of
distribution of site energies, namely, the Gaussian and
the binary potential distributions \cite{FABHRSWW,Markos}.
The exponents
at the critical points
are estimated to be $\alpha = 0.85 \pm 0.02$
for the Gaussian distribution
and $\alpha = 0.85 \pm 0.03$ for the binary distribution.
Together with the result for the uniform distribution,
the present results
suggest the universal fractal dimensionality at the critical
points of two-dimensional symplectic systems.

We have also performed the simulations in the metallic
regime for $W=3,4$ and $5V$. Here the auto-correlation
function keeps to show the power-law decay. The estimated
values of the exponent $\alpha$ are summarized in Table 1.
As the strength of disorder approaches to the critical
value $W_c$, the exponent starts to deviate from 1.
This crossover phenomenon might be interpreted as
a finite-size effect. As
demonstrated in ref.\cite{SG}, a fractal structure can be
observed at length-scales much smaller than the characteristic
length of the system.
The characteristic length of the metallic regime, namely the
correlation length,
becomes larger than the size of the system
as the critical point is approached and diverges
at the critical point
\cite{LR}.
It is thus expected that the exponent $\alpha$ deviates from 1
accordingly near the critical point,
reflecting the fractal character
of the present system at length-scales
much smaller than the characteristic length.

Let us consider, next, the behavior of the second  moment
of the wave packets. Its growth in the metallic regime for
$W=1,3,4$ and $5V$ are shown in Fig. 2. Here,
the second moment increases linearly in time
and thus, within the present length-scales,
the conductivity stays at a finite value.
This linear increase of second moment is also observed
at the critical point as shown in Fig. 3.

{}From our estimates of the diffusion coefficient and
the density of states per area $\rho \approx 0.25/(a^2V)$,
the conductivity at the critical point $(W=W_c)$ can be
evaluated as
$\sigma_c \approx 1.5(e^2/h)$, which is consistent with the
estimation in ref. \cite{Markos}.
In the localized regime, we observe that
the growth of the second moment
is apparently suppressed (Fig. 3).
The values of the diffusion coefficient $D$ estimated
using the relation (\ref{3})
for
various values of $W$ in the metallic regime
are also listed in Table 1.

Fal'ko and Efetov have recently proposed a formula for
the fractal dimensionality $D_2$
in terms of the diffusion coefficient
$D$ in the weakly
localized regime $(2\pi\rho D \hbar \gg 1)$\cite{FE}.
In the case of the symplectic ensemble,
it is expressed as  $D_2 = 2 - 1/(2 \pi^2 \rho D)$.
Together with $\alpha = D_2 /2 $,
our results (Table 1) are in good agreement with this formula.

In summary, we have performed numerical simulation for
the diffusion of electrons in the disordered
two-dimensional system
with the spin-orbit interaction. The present
method
preserves
the time-reversal symmetry which characterizes
the symplectic universality class.
In the
metallic regime away from the critical point,
the asymptotic behaviors of the temporal
auto-correlation function and of the second moment of the
wave packets are nothing but those expected from the
diffusion equation.
Within our simulation, there is no evidence
for the logarithmic divergence
of the conductivity, which is a consequence of a
direct application of the scaling theory\cite{HLN,Kawabata}.
At the critical point, the
temporal auto-correlation function shows the power-law decay with
a non-conventional exponent reflecting
the fractal structure in energy spectrum and in the wave function,
while the growth of the second moment seems to be diffusive.
We note that this type of behavior has been
observed in the critical states in the quantum Hall system\cite{HS}
as well as in the system with random magnetic fields\cite{KO}.
It would therefore represent a common property at the
critical point of the Anderson transition in two dimensional systems.
The
value of the exponent $\alpha$
at the critical point is consistent with the correlation dimension
$\tilde{D}_2$ obtained in ref. \cite{Schweitzer} and
the relation $\alpha = \tilde{D}_2$ \cite{KPG}.
We have further evaluated
the exponent $\alpha$ at the critical points
for several types of the potential
distributions and found that those values coincide with each other
within the present numerical accuracy.
This fact indicates the universality of the
exponent $\alpha$ at the critical points in two-dimensional
symplectic systems, which would enable us to determine
the critical points by using the equation of motion method.

The authors thank Y. Ono and M. Takahashi for useful discussions.
They also thank V.I. Fal'ko for
sending us the preprint (ref.\cite{FE})
and for communications.
Numerical calculations were done on HITAC S-3800 of the Computer
Center, University of Tokyo. This work was supported by the
Grant-in-Aid No.06740316 from the Ministry of Education,
Science and Culture, Japan. One of the authors (T.K.) thanks
the Fujukai Foundation for  financial support.


%
%
\begin{figure}
\caption{The temporal auto-correlation function $C(t)$
for the extended regime $(W=1V)$, the localized regime $(W=9V)$ and
at the critical point $(W=W_c)$. The exponent $\alpha$ is estimated
as $0.84 \pm 0.03$ and $0.996 \pm 0.014$ for $W=W_c$ and $W=1V$,
respectively.}
\end{figure}
\begin{figure}
\caption{The growth of the second moment in the metallic regime for
$W=1,3,4$ and $5V$. Solid lines correspond to $D=17.2$,
$2.87$, $1.99$ and $1.45 a^2 V/\hbar$ for $W=1$, $3$,
$4$, and $5V$, respectively.}
\end{figure}
\begin{figure}
\caption{The growth of the second moment at the critical point
($W=W_c$) and in
the localized regime ($W=8$, $9V$). Solid line for
$W=W_c$ corresponds
to $D=0.97 a^2V/\hbar$.}
\end{figure}
\begin{table}
\begin{tabular}{llll}
$W/V$ & $D/(a^2 V /\hbar)$ &$\rho/(1/(a^2V))$& $\alpha$ \\ \hline
1 & $\sim$ 17.2 &$\sim$ 0.36&$0.996 \pm 0.014$ \\
3 & $\sim$ 2.87 &$\sim$ 0.35&$0.97 \pm 0.04$ \\
4 & $\sim$ 1.99 &$\sim$ 0.30&$0.96 \pm 0.03$ \\
5 & $\sim$ 1.45 &$\sim$ 0.27&$0.93 \pm 0.03$ \\
$5.74(=W_c/V)$ & $\sim$ 0.97&$\sim$ 0.25&$0.84 \pm 0.03$
\end{tabular}
\caption{List of the exponent $\alpha$,
the diffusion coefficient $D$
and the density of states $\rho$
for the disorder $W/V = 1,3,4,5$ and $5.74(=W_c/V)$.
Here $a$ denotes the
lattice spacing.}
\end{table}
\end{document}